\documentclass[preprint,showpacs,preprintnumbers,nofootinbib,superscriptaddress]{revtex4}
\usepackage{graphicx}
\usepackage{epsfig}
\usepackage{dcolumn}
\usepackage{bm}

\def\gsim{\lower0.5ex\hbox{$\:\buildrel >\over\sim\:$}}
\def\lsim{\lower0.5ex\hbox{$\:\buildrel <\over\sim\:$}}

\newcommand{\bea}{\begin{eqnarray}}
\newcommand{\eea}{\end{eqnarray}}

\def\mpT{p_T \hspace{-.9em}/\;\:}

\def\stau{{\widetilde\tau}}
\def\smu{{\widetilde\mu}}
\def\se{{\widetilde e}}
\def\identity{\providecommand{\openone}{\leavevmode\hbox{\small1\kern-3.8pt\normalsize1}}}

\begin{document}

\preprint{HIP-2006-24/TH}
\preprint{IFT-UAM/CSIC-06-25}

\title{Lepton flavour violation in future linear colliders \\ in the long-lived 
stau NLSP scenario}

\author{Alejandro Ibarra}
\email{ibarra@mail.cern.ch}
\affiliation{Instituto de F\'{i}sica Te\'{o}rica, CSIC/UAM, C-XVI \\
Universidad Aut\'{o}noma de Madrid, \\
Cantoblanco, 28049 Madrid, Spain} 
\author{Sourov Roy}
\email{roy@cc.helsinki.fi}
\affiliation{Helsinki Institute of Physics, P.O. Box 64, FIN-00014 
University of Helsinki, Finland}



\begin{abstract} \vspace*{10pt}
We analyze the prospects of observing lepton flavour violation 
in future $e^- e^-$ and $e^+ e^-$ linear colliders in scenarios where 
the gravitino is the  lightest supersymmetric particle, and the 
stau is the next-to-lightest supersymmetric particle. The signals consist of
multilepton final states with two heavily ionizing charged tracks produced by the
long-lived staus. The Standard Model backgrounds are very small
 and the supersymmetric backgrounds can be kept well under 
control by the use of suitable kinematical cuts. 
We discuss in particular the potential of the projected
International Linear Collider to discover lepton flavour
violation in this class of scenarios, and we compare 
the estimated sensitivity with the constraints stemming 
from the non-observation of rare decays.
\end{abstract}

\maketitle

\section{Introduction}

The discovery of flavour violation in neutrino oscillations \cite{evidences}
has opened up a new era for flavour physics in the leptonic sector.
This crucial discovery also encourages the search for flavour
violation in the charged lepton sector, and if low energy supersymmetry 
is discovered, in the slepton sector. 

The scalar sector of supersymmetric theories contains many
new flavour violating couplings, stemming from the off-diagonal 
elements of the soft breaking terms. Namely, in the
mass eigenstate basis for the charged leptons, the soft-breaking
Lagrangian reads:
\bea
-{\cal L}_{\rm soft}=
(m^2_{\widetilde l_L})_{ij} {{\widetilde l_{L_i}}^{\dagger}} 
{\widetilde l_{L_j}} +
(m^2_{\widetilde l_R})_{ij} {{\widetilde l_{R_i}}^{\dagger}} 
{\widetilde l_{R_j}} +
(Y^l_{i} {A_l}_{ij} {\widetilde l_{R_i}}{\widetilde l_{L_j}} H_d  +h.c.),
\label{lagrangian}
\eea
where $i,j=1,2,3$ are generational indices. ${\widetilde l_{L,R}}$ denote
the left and right-handed charged sleptons, 
$H_d$ is the down-type Higgs doublet,
$Y^l_{i}$ is the charged-lepton Yukawa coupling, 
and $m^2_{\widetilde l_{L,R}}$ and $A_l$ are the soft
scalar mass matrix squared and the soft trilinear matrix, respectively.
The resulting $6\times 6$ charged-slepton mass matrix reads:
\bea
{\cal M}^2=
\pmatrix{m^2_{\widetilde l_L} + m^2_{l} 
-(\frac{1}{2}-\sin^2 \theta_W)m^2_Z \cos 2\beta &
m_l(A_l-\mu \tan \beta)^{\dagger} \cr
m_l(A_l-\mu \tan \beta) &
m^2_{\widetilde l_R} + m^2_{l} -\sin^2 \theta_W m^2_Z \cos 2\beta},
\eea
with $m_l=Y^l\langle H^0_d \rangle$. It is customary to express this 
mass matrix in the form
\cite{Hall:1985dx}:
\bea
{\cal M}^2=
\pmatrix{{m^{av}_{L}}^2({\bf 1}+\delta_{LL})&
m_l(A^{av}_l-\mu \tan \beta)^* + m^{av}_L m^{av}_R \delta_{LR}\cr
m_l(A^{av}_l-\mu \tan \beta) + m^{av}_L m^{av}_R \delta_{RL} &
{m^{av}_{R}}^2 ({\bf 1}+\delta_{RR})},
\label{mass-matrix2}
\eea
so that the amount of flavour violation is parametrized
by the $3\times3$ matrices  $\delta_{LL}$, $\delta_{RR}$,
 $\delta_{LR}$ and  $\delta_{RL}$, and we assume them to be real in this 
paper. In this expression, $m^{av}_{L}$, $m^{av}_{R}$ are 
an average of the masses of the left-handed 
and right-handed sleptons, respectively, and
$A^{av}_l$ is an average of the diagonal elements of
the soft trilinear matrix.

The size of these flavour violating terms
is model dependent, although very rarely strictly vanishing \cite{offdiag}.
They arise when the mechanism of supersymmetry breaking
distinguishes among flavours, thus producing flavour violation
in the soft breaking terms already at tree level. For example, 
in the framework of the Froggatt-Nielsen mechanism 
\cite{Froggatt:1978nt}, the flavon fields not
only generate the mass hierarchies and the mixings, 
but in general contribute
to the  breaking of supersymmetry, inducing off-diagonal
elements in the soft terms \cite{Ross:2002mr}. 
Besides, in weakly coupled string constructions where the K\"ahler
moduli fields participate in the breaking of supersymmetry,
if the matter metric is non-diagonal also flavour violating
couplings in the scalar sector are generated  
\cite{Kaplunovsky:1993rd,Brignole:1993dj}. Furthermore, even if the
supersymmetry breaking mechanism is flavour blind, 
so that the soft terms are flavour diagonal at the cut-off scale, 
radiative effects might spoil this diagonal structure.
Many well motivated models predict the existence of particles
in the desert with flavour violating couplings to the
Minimal Supersymmetric Standard Model (MSSM) superfields, 
that would generate off-diagonal entries in
the soft-breaking matrices through quantum corrections.
Renowned examples are the MSSM extended with right-handed
neutrinos \cite{Borzumati:1986qx} and Grand Unified models 
\cite{Barbieri:1994pv}.

Flavour violation in the scalar sector is propagated
through loop effects to the charged lepton sector,
inducing rare decays. The non-observation of these
processes imposes very strong constraints on the 
flavour violating couplings, that depend on the
particular point of the SUSY parameter space. 
For future reference, we show in table \ref{Table1} the
constraints on the flavour violating parameters 
$\delta_{LL}$, $\delta_{RR}$, $\delta_{LR}$ and  $\delta_{RL}$
for the supersymmetric spectrum of the benchmark
point $\epsilon$ of \cite{DeRoeck:2005bw} that
follow from the present experimental bounds 
$BR(\mu\rightarrow e\gamma) \lsim1.2\times10^{-11}$ \cite{Brooks:1999pu},
$BR(\tau\rightarrow \mu\gamma) < 3.1\times 10^{-7}$ \cite{Abe:2003sx}
(Belle) or
$<6.8\times10^{-8}$ \cite{Aubert:2005ye} (BaBar), and 
$BR(\tau\rightarrow e\gamma) <3.9\times 10^{-7}$ \cite{Hayasaka:2005xw}.
The relevant formulae to compute the bounds on the
$\delta$'s from the experimental constraints on the
rare decays can be found in the Appendix.
The $\epsilon$ benchmark point 
belongs to the class of scenarios that we would
like to study in this paper, namely when the gravitino is
the lightest supersymmetric particle and the stau is
the next-to-lightest supersymmetric particle, and will be
used in this paper to illustrate our results. The complete
Higgs and supersymmetric spectra for this benchmark point
is summarized in table \ref{Table2}.

The next round of
experiments will search for muon and tau rare decays
with enhanced sensitivity \cite{flavour-experiments},
and will provide improved bounds on the flavour 
violating couplings, or hopefully, a positive signal 
for lepton flavour violation. On the other hand, the advent
of high energy particle colliders offers new opportunities
to study flavour violation. The on-shell production
of supersymmetric particles would allow the study of
their tree-level flavour violating production and decay. 
This strategy would offer very valuable information about
the scalar sector, that would complement the information
provided by rare decays.

\begin{table}
\begin{center}
\begin{tabular}{|cc|cc|cc|cc|c|} \hline
 sector && $\delta_{LL}$ && $\delta_{RR}$ && $\delta_{LR}$ && $\delta_{RL}$
\\ \hline \hline
12 && $2\times10^{-4}$    &&  $6\times10^{-4}$ &&
$4\times10^{-6}$    &&     $4\times10^{-6}$
\\ \hline 
13 &&  0.09   &&  0.27   && 0.03  && 0.03     
\\ \hline 
23 &&  0.04   &&  0.11   && 0.01   &&  0.01   
\\ \hline 
\end{tabular}
\end{center}
\caption{\protect\footnotesize Constraints on the 
flavour violating parameters $\delta_{LL}$, $\delta_{RR}$,
 $\delta_{LR}$ and  $\delta_{RL}$, for
the supersymmetric spectrum of the benchmark point 
$\epsilon$ of \cite{DeRoeck:2005bw} (see details in the text).}
\label{Table1}
\end{table}

\begin{table}
\begin{center}
\begin{tabular}{|cc cc||cc cc||cc cc||cc cc |}
 \hline 
$h^0$ && 119  && $\chi^0_1$ && 183 && 
$\widetilde e_L$, $\widetilde \mu_L$ && 298 &&
$\widetilde u_L$, $\widetilde c_L$ && 897 &
\\
$H^0$ && 641 && $\chi^0_2$ && 349&& 
$\widetilde e_R$, $\widetilde \mu_R$ && 169 &&
$\widetilde u_R$, $\widetilde c_R$ && 867&
\\
$A^0$ && 641 && $\chi^0_3$ && 578&& 
$\widetilde \nu_e$, $\widetilde \nu_\mu$ &&287 &&
$\widetilde d_L$, $\widetilde s_L$ && 901&
\\
$H^{\pm}$ && 646&& $\chi^0_4$ && 593&& 
$\widetilde \tau_1$ && 150&&
$\widetilde d_R$, $\widetilde s_R$ && 864&
\\
          &&    && $\chi^{\pm}_1$ && 349&& 
$\widetilde \tau_2$ && 302&&
$\widetilde t_1$ && 682&
\\
          &&    && $\chi^{\pm}_2$ && 594&& 
$\widetilde \nu_\tau$ && 285&&
$\widetilde t_2$ && 879 &\\
          &&    && $\widetilde g$ && 986 &&  && &&
$\widetilde b_1$ && 824 &\\
          &&    &&  && &&  && &&
$\widetilde b_2$ && 862 &
\\ \hline 
\end{tabular}
\end{center}
\caption{\protect\footnotesize Mass spectrum (in GeV) for 
the benchmark point $\epsilon$, taken from Table 2
in \cite{DeRoeck:2005bw}.}
\label{Table2}
\end{table}

The absence of large hadronic backgrounds at
$e^+ e^-$ or $e^- e^-$ colliders makes this class
of experiments particularly convenient to study
lepton flavour violation \cite{Aguilar-Saavedra:2001rg}.
The signals will depend crucially on the particular 
supersymmetric scenario considered. We can differentiate
roughly two main classes of scenarios, according to the nature
of the lightest supersymmetric particle (LSP). A popular
choice for the LSP is the neutralino, although 
scenarios with superweakly interacting LSP, such as the gravitino or the 
axino, are also compatible with all the collider experiments
and cosmology. In this paper we will concentrate on the latter
class of scenarios, focusing for definiteness on the case
with gravitino LSP, although the analysis and the
conclusions for the case with axino LSP are completely analogous.
These scenarios have received
a lot of attention recently and their properties have been
studied in a number of papers 
\cite{DeRoeck:2005bw,stopped-staus,superWIMP,Hamaguchi:2004ne}.
One of the most remarkable features of these scenarios is the
longevity of the next-to-lightest supersymmetric particle (NLSP).
The NLSP could only decay to gravitinos
and Standard Model particles through gravitational interactions,
with a decay rate strongly suppressed by the Planck mass.
This translates into lifetimes that could be long enough
to allow the NLSP to traverse the detector.

Under the assumption of universality of the soft
breaking scalar, gaugino and trilinear parameters at
some cut-off scale,
the NSLP can be either a neutralino or a stau, although
in more general scenarios other candidates could also be possible.
In the case that the NLSP is the neutralino, the signals for
lepton flavour violation will be identical to the case with
neutralino LSP, which have been discussed extensively in the
literature \cite{LFV-neutralino}. On the other hand, the signals for the case
with stau NLSP could be very different. Whereas the neutralino
escapes detection and is identified as missing energy, the stau
would produce a heavily ionizing track in the vertex  detector.
This signature is very unique, therefore the observation of
heavily ionizing tracks would give strong support to this
scenario and would allow the search for lepton flavour
violation essentially without Standard Model backgrounds.

The longevity of the staus
would also allow their collection and the subsequent detection of their decay
products with reduced backgrounds. This possibility has been 
discussed in \cite{DeRoeck:2005bw,stopped-staus}, and in particular
the possibility of detecting lepton flavour violation was 
studied in \cite{Hamaguchi:2004ne}.

This paper is organized as follows. In Section II, we present
the notation and the setup for our analysis, and discuss
the different signatures for lepton flavour violation at
future $e^- e^-$ and $e^+ e^-$ colliders. In Section III,
we describe in particular the analysis for the case of the projected
International Linear Collider (ILC) and estimate the
sensitivity reach of this experiment to lepton flavour
violation in scenarios where the gravitino is the LSP and
the stau the NLSP. We also compare this sensitivity
with the present and future constraints on lepton flavour violation
stemming from the non-observation of rare leptonic decays.
In Section IV we present our conclusions, and finally, in
the Appendix, we include the relevant formulas to
compute the branching ratios for the process 
$\ell_i\rightarrow \ell_j \gamma$.

\section{Flavour violating signatures at future electron colliders}

Throughout this paper we will assume that the gravitino is the LSP
and the NLSP is mainly a right-handed stau, although it could
have some admixture of left-handed stau or other leptonic
flavours. We will denote the mass eigenstates by the dominant
flavour, so that the NLSP will be denoted by ${\stau_1}$. 
Motivated by the low energy spectrum of the constrained MSSM, we
will also assume that next to the ${\stau_1}$, the lightest superparticles 
are the two combinations of right-handed selectron and smuon, also
with a very small admixture of left-handed states (due to the Yukawa
suppression of the left-right mixing) and some admixture
of stau. The mass splitting between them is expected to be very small,
and the absolute values of their masses are expected to be not very different 
from the NLSP mass. We will denote these states by $\se_R$ and 
$\smu_R$, the former being the mass eigenstate with largest
right-handed selectron component and the latter with largest
right-handed smuon component. Next in mass in the supersymmetric
spectrum are the lightest neutralino and the rest of the sparticles.
Schematically the spectrum reads
\bea
m_{3/2}<m_{\stau_1}<m_{\se_R,\;\smu_R}<m_{\chi^0_1}, 
m_{\se_L,\smu_L}, m_{\stau_2}...
\eea

It is important for our analysis that the NLSP decays outside
the detector, so that it is detected as a heavily ionizing track.
If R-parity is conserved, the NLSP can only decay  
into a gravitino and a charged lepton with total decay rate
\bea
\Gamma
&\simeq& 
\frac{m^5_{\stau_1}}{48 \pi m^2_{3/2} M^2_P}
\left(1-\frac{m^2_{3/2}}{m^2_{\stau_1}}\right)^4,
\label{decay}
\eea
where $M_P= (8 \pi G_N)^{-1/2}$ is the reduced Planck mass.
Therefore, the requirement that the NLSP decay length is 
larger than ten centimeters, to guarantee that the NLSP traverses
a few layers in the vertex detector leaving a  heavily ionizing track,
translates into the constraint on the gravitino
mass $m_{3/2}\gg 0.1~\mathrm{keV}\times(m_{\stau}/100~\mathrm{GeV})^{5/2}$.

After discussing the set-up for this analysis, let us discuss
the possible signals and backgrounds for the detection
of lepton flavour violation. Despite the fact that the 
discussion is very similar
for the $e^- e^-$ and for the $e^+e^-$ collider, let us analyze,
for the sake of clarity, each case separately. 

\subsection{$e^- e^-$ collider}

When lepton flavour is conserved, only left and right-handed selectrons 
would be produced in this mode, to be precise in the $t$-channel by 
neutralino exchange. Since right-handed selectrons are lighter than left-handed
selectrons, the largest  production cross section would correspond to the 
process $e^-\; e^-\rightarrow\se^-_R\; \se^-_R$. The signatures for this 
process depend crucially on the mass splitting between the right-handed 
selectron and the NLSP. If the mass splitting is sufficiently large 
($\sim$ 15 -- 20 GeV), the right-handed selectron would decay mainly into charged 
leptons and a NLSP before reaching the detector, in a process mediated by 
neutralinos. The signature for this lepton flavour conserving process would be 
the detection of two heavily ionizing tracks, two electrons and two taus
\footnote{To be precise,in the final state one could find 
two taus, two antitaus or a pair of tau-antitau, depending on the 
charges of the outgoing NLSPs. However, and for simplicity in 
the notation, we will denote as ``tau'' either the
tau or the antitau, and analogously for the electron and the muon.}
\bea
e^-\; e^- \rightarrow \se^-_R\; \se^-_R
\rightarrow (e^- \; \tau^{\pm} \; \stau^{\mp}_1) (e^-\; 
\tau^{\pm} \; \stau^{\mp}_1).
\eea
If the mass splitting between the right-handed selectron and the 
NLSP is smaller, the charged leptons could be too soft to be detected, 
and only the two heavily ionizing tracks would be observed. 
Finally, if the mass splitting is very small, the decay channel 
into charged leptons would be kinematically closed. 
Selectrons could only decay into  NLSPs and neutrinos by a process mediated by 
charginos, with a decay rate suppressed by the small electron Yukawa coupling:
\bea
e^-\; e^- \rightarrow \se^-_R\; \se^-_R
\rightarrow (\nu_e \; {\bar\nu_{\tau}} \; \stau^{-}_1) 
(\nu_e \; {\bar\nu_{\tau}}  \; \stau^{-}_1) .
\eea
If this is the case, again only two heavily ionizing tracks
would be detected, corresponding to the NLSPs or perhaps to the
right-handed selectrons, if these are long lived enough to traverse
the detector. Therefore, production of two right-handed
selectrons would result in the detection of two heavily 
ionizing tracks, two electrons and two taus; or the detection
of just two heavily ionizing tracks.

Less likely than the production of two right-handed selectrons
is the associated production of one left-handed selectron and one
right-handed selectron. Left-handed selectrons can decay
via neutralino exchange either into charged leptons and a NLSP,
or via chargino exchange into neutrinos and a NLSP, with comparable decay 
rates.  Therefore, in the detector two heavily ionizing tracks would
be observed, with either two electrons and two taus, or one electron and 
one tau, or no charged leptons:
\bea
e^-\; e^- &&\rightarrow \se^-_L\; \se^-_R\rightarrow
(e^- \; \tau^{\pm} \; \stau^{\mp}_1) 
(e^- \; \tau^{\pm} \; \stau^{\mp}_1), \label{emem-seLseR-4cl} \\
e^-\; e^- &&\rightarrow \se^-_L\; \se^-_R
\rightarrow (\nu_e \; {\bar\nu_{\tau}} \; \stau^{-}_1) 
(e^- \; \tau^{\pm} \; \stau^{\mp}_1), \label{emem-seLseR-2cl} \\
e^-\; e^- &&\rightarrow \se^-_L\; \se^-_R
\rightarrow (\nu_e \; {\bar\nu_{\tau}} \; \stau^{-}_1) 
(\nu_e \; {\bar\nu_{\tau}}  \; \stau^{-}_1).\label{emem-seLseR-0cl} 
\eea
(The first two processes would have comparable cross
sections, whereas the third would be suppressed by the 
small electron  Yukawa coupling in the decay of $\se^-_R$.)
Similar signatures would follow from the production of two
left-handed selectrons, in this case with comparable cross 
sections for the three processes. As we will see, processes 
associated with the production of left-handed selectrons 
constitute the most important source of background for
the detection of lepton flavour violation.

If lepton flavour violation exists in nature at an 
observable level, novel possibilities arise, namely the
associated production of a right-handed selectron 
and a smuon, or a right-handed selectron and a NLSP. 
The associated production
of a right-handed smuon and a NLSP could also be possible, 
although it would require two flavour violating vertices 
and is therefore very suppressed.

The associated production
of $\se^-_R$ and $\smu^-_R$ would give rise to the observation
of two heavily ionizing tracks, two taus, one muon and one
electron:
\bea
e^-\; e^- &&\rightarrow \se^-_R\; \smu^-_R\rightarrow
(e^- \; \tau^{\pm} \; \stau^{\mp}_1) 
(\mu^- \; \tau^{\pm} \; \stau^{\mp}_1).
\label{emem-seRsmuR-mue2tau}
\eea
No other process in an $e^-e^-$ collider yields
this same signal, and if particle identification
is sufficiently good, the observation of this
process would represent a clear signal for 
lepton flavour violation in the  selectron-smuon sector,
parametrized by the off-diagonal element of the right-handed
slepton mass matrix $(m^2_{\widetilde l_R})_{12}$.

On the other hand, the associated production of $\se^-_R$ and
$\stau^-_1$ would give rise to the observation
of two heavily ionizing tracks
(when the mass splitting between the right-handed selectron
and the NLSP is small), or two heavily ionizing tracks,
one electron and one tau (when the mass splitting is sufficiently large).
The former process cannot be distinguished from the production
of two right-handed selectrons and cannot be used for
the search of lepton flavour violation. However, the
latter process could constitute a strong signal of lepton flavour
violation:
\bea
e^-\; e^- \rightarrow \se^-_R\; \stau^-_1
\rightarrow (e^- \; \tau^{\pm} \; \stau^{\mp}_1)\;  \stau^-_1.
\label{emem-seRstau1-2cl}
\eea

The final state in this processes could be mimicked by the
production of two right-handed selectrons, and the subsequent decay
of one of them into neutrinos and a NLSP by an interaction mediated
by higgsinos, while the other decays into charged leptons
and a NLSP: 
$e^-e^-
\rightarrow {\tilde e}^-_R {\tilde e}^-_R \rightarrow (e^- \tau^+ {\tilde \tau}^-_1 {\tilde
\tau}^-_1 \nu_e {\bar \nu}_\tau + e^- \tau^- {\tilde \tau}^+_1 {\tilde \tau}^-_1 \nu_e
{\bar \nu}_\tau$). However, the
decay mode ${\tilde e}_R \rightarrow \nu_e {\bar \nu}_\tau {\tilde \tau}^-_1$ is
highly suppressed due to the presence of the electron Yukawa coupling,
which implies decay rates of the order of $10^{-19}$ GeV. 
Hence, this source of background can  be neglected in general.

A more important background could follow from the associated production
of a left-handed and a right-handed selectron, or two left-handed selectrons, 
where the left-handed selectron decays into neutrinos and a NLSP with {\em unsuppressed}
decay rate, Eq.(\ref{emem-seLseR-2cl}). One should note here that these 
background events include two neutrinos in the final state which give rise to 
imbalance in momentum. On the other hand, the signal events, 
Eq.(\ref{emem-seRstau1-2cl}), do not have any neutrinos in the final state 
and hence there is no missing transverse 
momentum $\mpT$. Therefore, if we demand that the final state should have a 
$\mpT <$ 5 GeV then most of the background could be eliminated. This 
background could be further reduced using right-polarized 
electron beams. 
In addition, the signal cross section increases with 
the use of right-polarized beams. Nevertheless, it should be kept 
in mind that the luminosity is also reduced when one considers polarized 
beams since only a part of the total integrated luminosity available is used 
for polarization. In consequence, the signal significance (defined later) 
in the case of right-polarized beams does not change much compared to that 
in the case of unpolarized beams and hence the sensitivity of the experiment
remains almost unchanged in both cases.
Another way to eliminate this background (with a previous 
knowledge of the spectrum) is to tune the center of mass energy of the 
collider to be below the threshold for left-handed selectron production. 

Additional signals for lepton flavour violation in the right-handed
sector follow from the lepton flavour violating decays of the right-handed selectrons.
Depending on which particular sector violates lepton flavour, right-handed
selectrons could decay into two electrons 
and a NLSP or two taus  and a NLSP, 
if the violation occurs in the right-handed 
selectron-stau sector, {\it i.e.} $ (m^2_{\widetilde l_R})_{13}\neq 0$; 
one muon, one tau  and a NLSP if it occurs in the
right-handed selectron-smuon sector, {\it i.e.} $(m^2_{\widetilde l_R})_{12}\neq 0$;
or one electron, one muon  and a NLSP if it occurs in the right-handed 
smuon-stau sector, {\it i.e.} $(m^2_{\widetilde l_R})_{23}\neq 0$. The 
corresponding lepton flavour violating processes read:
\bea
e^-\; e^-&\rightarrow \se^-_R\; \se^-_R \rightarrow 
(e^- \; e^{\pm} \; \stau^{\mp}_1)\; 
(e^- \; \tau^{\pm} \; \stau^{\mp}_1) ~~~~ {\rm if}~~ 
(m^2_{\widetilde l_R})_{13}\neq 0,
\label{emem-seRseR-3etau} \\
e^-\; e^-&\rightarrow \se^-_R\; \se^-_R \rightarrow 
(\tau^- \; \tau^{\pm} \; \stau^{\mp}_1)\; 
(e^- \; \tau^{\pm} \; \stau^{\mp}_1)~~~~ {\rm if}~~
 (m^2_{\widetilde l_R})_{13}\neq 0 ,
\label{emem-seRseR-e3tau}\\
e^-\; e^-&\rightarrow \se^-_R\; \se^-_R \rightarrow 
(\mu^- \; \tau^{\pm} \; \stau^{\mp}_1)\; 
(e^- \; \tau^{\pm} \; \stau^{\mp}_1)~~~~ {\rm if}~~ 
(m^2_{\widetilde l_R})_{12}\neq 0 ,
\label{emem-seRseR-mue2tau}\\
e^-\; e^-&\rightarrow \se^-_R\; \se^-_R \rightarrow 
(e^- \; \mu^{\pm} \; \stau^{\mp}_1)\; 
(e^- \; \tau^{\pm} \; \stau^{\mp}_1)~~~~ {\rm if}~~ 
(m^2_{\widetilde l_R})_{23}\neq 0.
\label{emem-seRseR-mu2etau}
\eea
None of these processes suffers from important backgrounds,
provided particle identification is sufficiently reliable,
and constitute important probes of lepton flavour violation
in the right-handed slepton sector.

\subsection{$e^+\; e^-$ collider}

Production of sleptons at the $e^+ e^-$ collider proceeds via 
$t$-channel by neutralino exchange and via $s$-channel
by photon and $Z$ boson exchange.
In the $t$-channel the possible processes are analogous to those for the
$e^-e^-$ collider, with the appropriate changes in the electric charges of
the particles. Namely, when lepton flavour is conserved, only selectrons
will be produced, and when lepton flavour is violated, the analogous
processes to Eqs.(\ref{emem-seRsmuR-mue2tau})-(\ref{emem-seRseR-mu2etau})
will occur, with analogous backgrounds.

On the other hand, in the $s$-channel all types of sleptons can be 
produced: 
\bea
e^+\; e^-\rightarrow {\widetilde l_{R_i}}^+\; {\widetilde l_{R_j}}^-, \\
e^+\; e^-\rightarrow  {\widetilde l_{R_i}}^{\pm}\; 
{\widetilde l_{L_j}}^{\mp}, \\
e^+\; e^-\rightarrow {\widetilde l_{L_i}}^+\; {\widetilde l_{L_j}}^-,
\eea 
where ${\widetilde l}_i,\; {\widetilde l}_j = \se,\; \smu,\; \stau$.

If lepton flavour is conserved, sleptons would be produced
in pairs with opposite lepton family number. The production
of a pair of NLSPs with opposite charges would be detected as
two back-to-back heavily ionized tracks
whereas production of pairs of right-handed 
smuons or selectrons would be detected
as two heavily ionized tracks, together with two taus and two muons 
or electrons, respectively. Right-handed  smuons and  selectrons
could also decay into neutrinos via higgsino exchange, yielding
a signature consisting of just two heavily ionizing tracks plus
missing energy. Nevertheless, these processes are very
suppressed by the small electron and muon Yukawa couplings
and can be usually neglected,
even in the large $\tan\beta$ regime.

The experimental signatures are qualitatively different 
when left-handed sfermions are produced.
These could decay via gaugino
exchange either into charged leptons or into neutrinos
with comparable decay rates,  yielding signatures with
two heavily ionizing tracks and four, 
two or no charged leptons in the final state.
For the case of selectron production, the possible final 
states are
\bea
e^+\; e^-&\rightarrow& \se^+_R\; \se^-_L \rightarrow 
(e^+ \;\tau^{\pm} \;\stau^{\mp}_1)\;
(e^- \;\tau^{\pm} \;\stau^{\mp}_1),\\
\label{e+e-lfvprod-bkgd}
e^+\; e^-&\rightarrow& \se^+_R\; \se^-_L \rightarrow 
(e^+ \;\tau^{\pm} \;\stau^{\mp}_1)\;
(\nu_{e}\; \bar \nu_{\tau} \; \stau^{-}_1),\\
e^+\; e^-&\rightarrow& \se^+_L\; \se^-_L \rightarrow 
(e^+ \;\tau^{\pm} \;\stau^{\mp}_1)\;
(e^- \;\tau^{\pm} \;\stau^{\mp}_1),\\
e^+\; e^-&\rightarrow& \se^+_L\; \se^-_L \rightarrow 
(e^+ \;\tau^{\pm} \;\stau^{\mp}_1)\;
(\nu_{e}\; \bar \nu_{\tau} \; \stau^{-}_1),\\
e^+\; e^-&\rightarrow& \se^+_L\; \se^-_L \rightarrow 
(\bar \nu_{e}\;  \nu_{\tau} \; \stau^{+}_1)\;
(\nu_{e}\; \bar \nu_{\tau} \; \stau^{-}_1).
\eea
As for the case of the $e^-e^-$ linear collider, the left-handed
slepton decays into neutrinos will represent the most important 
source of background for the search of lepton flavour violation.

Lepton flavour violation could manifest itself either in the
production of sleptons or in their decays. Concentrating
just on lepton flavour violation in the right-handed 
slepton sector, the lepton flavour
violating photon or $Z$-boson vertices would give
rise to the following processes:
\bea
e^+\; e^-&\rightarrow&  \se^{+}_R \; \smu^{-}_R \rightarrow 
(e^+ \;\tau^{\pm} \;\stau^{\mp}_1)\;  (\mu^- \;\tau^{\pm} \;\stau^{\mp}_1)
 ~~~~ {\rm if}~~ (m^2_{\widetilde l_R})_{12}\neq 0 ,
\label{epem-smuRseR-4cl}\\
e^+\; e^-&\rightarrow& \smu^{+}_R\; \stau^{-}_1 \rightarrow 
(\mu^+ \;\tau^{\pm} \;\stau^{\mp}_1)\; \stau^{-}_1
 ~~~~ {\rm if}~~ (m^2_{\widetilde l_R})_{23}\neq 0 ,
\label{epem-smuRstau1-2cl}\\
e^+\; e^-&\rightarrow& \se^{+}_R\; \stau^{-}_1 \rightarrow 
(e^+ \;\tau^{\pm} \;\stau^{\mp}_1)\; \stau^{-}_1
 ~~~~ {\rm if}~~ (m^2_{\widetilde l_R})_{13}\neq 0.
\label{epem-seRstau1-2cl}
\eea
It is important to note that the lepton flavour violating 
photon and $Z$-boson vertices appear only at the one 
loop level. Consequently the cross sections
for these flavour violating processes in the $s$-channel are suppressed
compared to flavour violating processes in the $t$-channel,
and therefore will not be considered in our analysis.
Additional flavour violating signatures could stem
from the pair production of two neutralinos in the $s$-channel,
followed by the flavour violating decay of one of them into a charged lepton
and the NLSP, for example, $e^+\; e^-\rightarrow \chi^0 \chi^0 \rightarrow
(e^{\pm} \stau^{\mp}_1) (\tau^{\pm} \stau^{\mp}_1)$. The cross
section for this process is also smaller than that of the
flavour violating processes proceeding in the $t$-channel, 
since in this scenario neutralinos are heavier than sleptons, 
and will not be considered either.

Lepton flavour violation signals could also stem
from the decay of right-handed selectrons,
similarly to the case for the $e^-e^-$ collider, 
Eqs.(\ref{emem-seRseR-3etau}-\ref{emem-seRseR-mu2etau}):
\bea
e^+\; e^-&\rightarrow \se^+_R\; \se^-_R \rightarrow 
(e^+ \; e^{\pm} \; \stau^{\mp}_1)\; 
(e^- \; \tau^{\pm} \; \stau^{\mp}_1) ~~~~ {\rm if}~~ 
(m^2_{\widetilde l_R})_{13}\neq 0,
\label{epem-seRseR-3etau} \\
e^+\; e^-&\rightarrow \se^+_R\; \se^-_R \rightarrow 
(\tau^+ \; \tau^{\pm} \; \stau^{\mp}_1)\; 
(e^- \; \tau^{\pm} \; \stau^{\mp}_1)~~~~ {\rm if}~~ 
(m^2_{\widetilde l_R})_{13}\neq 0 ,
\label{epem-seRseR-e3tau}\\
e^+\; e^-&\rightarrow \se^+_R\; \se^-_R \rightarrow 
(\mu^+ \; \tau^{\pm} \; \stau^{\mp}_1)\; 
(e^- \; \tau^{\pm} \; \stau^{\mp}_1)~~~~ {\rm if}~~ 
(m^2_{\widetilde l_R})_{12}\neq 0 ,
\label{epem-seRseR-mue2tau}\\
e^+\; e^-&\rightarrow \se^+_R\; \se^-_R \rightarrow 
(e^+ \; \mu^{\pm} \; \stau^{\mp}_1)\; 
(e^- \; \tau^{\pm} \; \stau^{\mp}_1)~~~~ {\rm if}~~ 
(m^2_{\widetilde l_R})_{23}\neq 0,
\label{epem-seRseR-mu2etau}
\eea
and analogous processes when it is $\se^-_R$ the particle that
decays violating flavour instead of  $\se^+_R$. On the other
hand, production and lepton flavour violating decay 
of right-handed smuons would yield the following signals:
\bea
e^+\; e^-&\rightarrow \smu^+_R\; \smu^-_R \rightarrow 
(\mu^+ \; \mu^{\pm} \; \stau^{\mp}_1)\; 
(\mu^- \; \tau^{\pm} \; \stau^{\mp}_1) ~~~~ {\rm if}~~ 
(m^2_{\widetilde l_R})_{23}\neq 0,
\label{epem-smuRsmuR-3mutau} \\
e^+\; e^-&\rightarrow \smu^+_R\; \smu^-_R \rightarrow 
(\tau^+ \; \tau^{\pm} \; \stau^{\mp}_1)\; 
(\mu^- \; \tau^{\pm} \; \stau^{\mp}_1)~~~~ {\rm if}~~ 
(m^2_{\widetilde l_R})_{23}\neq 0 ,
\label{epem-smuRsmuR-mu3tau}\\
e^+\; e^-&\rightarrow \smu^+_R\; \smu^-_R \rightarrow 
(e^+ \; \tau^{\pm} \; \stau^{\mp}_1)\; 
(\mu^- \; \tau^{\pm} \; \stau^{\mp}_1)~~~~ {\rm if}~~ 
(m^2_{\widetilde l_R})_{12}\neq 0 ,
\label{epem-smuRsmuR-mue2tau}\\
e^+\; e^-&\rightarrow \smu^+_R\; \smu^-_R \rightarrow 
(\mu^+ \; e^{\pm} \; \stau^{\mp}_1)\; 
(\mu^- \; \tau^{\pm} \; \stau^{\mp}_1)~~~~ {\rm if}~~ 
(m^2_{\widetilde l_R})_{13}\neq 0,
\label{epem-smuRsmuR-e2mutau}
\eea
and analogously when $\smu^-_R$ decays violating flavour instead
of  $\smu^+_R$. No other process in the $e^+ e^-$ collider yields
the same signals, therefore, if particle identification is sufficiently
good, the observation of these processes would constitute robust
evidences for lepton flavour violation in the right-handed slepton sector.

\section{Discovery potential of the ILC}

In this section we analyze the prospects to observe lepton
flavour violation in the projected International Linear Collider (ILC),
both for the $e^- e^-$ mode and the $e^+e^-$ mode. 
For definiteness, we will assume a center of mass energy
$\sqrt{s}=500$ GeV, and an integrated luminosity of 500 fb$^{-1}$.
\footnote{The International Linear Collider is likely to operate
in two phases; the first with $\sqrt{s}=500$ GeV and the
second with $\sqrt{s}=1$ TeV, with an integrated luminosity
of 1ab$^{-1}$ for the $e^+e^-$ mode, and smaller for the 
$e^- e^-$ mode \cite{Kilian:2006he}.} We will also assume
in our numerical analyses that the beams are unpolarized, although 
as discussed in the previous section, the use of polarized beams
would enhance the strength of the lepton flavour violating signals and
reduce the background cross section. 
\footnote{The projected International Linear
Collider is expected to achieve the projected luminosity
of  500 fb$^{-1}$ with at least an  80\% electron polarization 
and a 60\% positron polarization at the interaction point.
A degree of polarization of a 90\% for the electrons and 
a 75\% for the positrons could possibly be achieved at the cost 
of some reduction in luminosity \cite{linearcollider}.}
The cross sections of the signal events have been calculated in the narrow 
width approximation. We have calculated the relevant 2$\rightarrow$2 
differential cross section ${\rm d}\sigma/{\rm dcos}\theta$ and then folded
into it the appropriate branching fractions of the corresponding decay 
channels to get the various final states described earlier.

We select the signal events according to the following criteria:
\begin{itemize}
\item{The transverse momentum of the electrons, the positrons and $\mu^\pm$ 
must be large enough: $p^{e^\pm,\mu^\pm}_T> 5$ GeV.}
\item{Slightly stronger selection criterion has been set on the transverse 
momentum of the $\tau$s: $p^{\tau}_T> 10$ GeV.}
\item{The transverse momentum of the ${\tilde \tau}_1$s must satisfy 
$p^{{\tilde \tau}_1}_T> 10$ GeV.}
\item{The electrons, the positrons, the $\mu^\pm$ and the $\tau$s and both 
the staus must be relatively  central, {\it i.e}. their pseudorapidities must 
fall in the range $|\eta^{{e^\mp},{\mu^\pm},{\tilde \tau}_1, {\tau}}|< 2.5$.}
\item{The electrons, the positrons, the $\mu^\pm$, the $\tau$s and 
the staus must be 
well-separated from each other: {\it i.e}. the isolation variable 
$\Delta R \equiv \sqrt{(\Delta \eta)^2 + (\Delta \phi)^2}$ (where $\eta$ and 
$\phi$ denote the separation in rapidity and the azimuthal angle, 
respectively) should satisfy $\Delta R> 0.4$ for each combination.}
\item{The missing transverse momentum $\mpT <$ 5 GeV.}
\end{itemize}

To illustrate the discovery potential of lepton flavour
violation in scenarios with stau NLSP at the ILC we will 
show contour plots of constant cross section for 
a variant of the $\epsilon$ benchmark point of \cite{DeRoeck:2005bw}. 
In our study we will take all the supersymmetric parameters as in 
the $\epsilon$ benchmark point, but we will vary the NLSP mass between 
144 GeV and 167 GeV (recall that in this benchmark point the mass of the 
next-to-NLSP, the right-handed selectron, is 169 GeV).
We will also admit some small amount of lepton flavour violation in
the right-handed slepton sector, parametrized by $\delta^{ij}_{RR}$.

At this stage it is very important to discuss the efficiency of
identification of the two approximately back-to-back stau tracks from 
muon tracks.
For sufficiently long-lived staus, 
there are two traditional ways of identification of slowly moving charged
massive particles$\colon$ (1) using a time-of-flight (ToF) device and
(2) measuring the associated high ionization energy loss rates $dE/dx$ in
the vertex detector and tracking chambers. In the case of a time-of-flight
device, one compares the mean time of flight for a muon ($\beta\simeq 1$)
to that of the long-lived massive charged particle ($\beta<1$) for a flight
path length of the order of a few meters. 
Considering a future linear collider operating with 
1.4 ns of bunch separation and a detector capable
to measure the time of flight with a 50 ps error,
it has been proved possible to identify the back-to-back tracks
as long-lived staus, with an efficiency varying between
$\sim$ 60--80$\%$ in the mass range relevant for our
scenario, 140--170 GeV \cite{Mercadante:2000hw}.

An alternative possibility to identify the staus is the measurement of
the rate of energy loss, $dE/dx$, that depends on the
$\beta \gamma$ of the particle. It has been shown in 
Ref.\cite{Mercadante:2000hw}, that assuming a 5\% resolution
in the measurement of $dE/dx$ and using suitable cuts,
it could be possible to identify the staus. However, 
the identification efficiency decreases significantly for masses around 
150 GeV, which are precisely the masses relevant for
our study\footnote{In this mass range the stau has $\beta \gamma\sim$ 1.3 
for a $\sqrt{s}$ = 500 GeV linear collider, and the energy deposited
coincides with the one by the muon (with  $\beta \gamma$ $\sim {\cal O} 
(10^3$)), hence the drastic reduction in the identification efficiency.}. 
Therefore, we will assume for our study that
the staus can be identified just with a ToF device and 
we will use for the efficiency a conservative value of a 60$\%$.
We will multiply the cross section of the signal events with
this efficiency in order to get realistic numbers. A more detailed analysis
on this issue is beyond the scope of this study.

\subsection{$e^-\; e^-$ collider}

\begin{figure}
\includegraphics{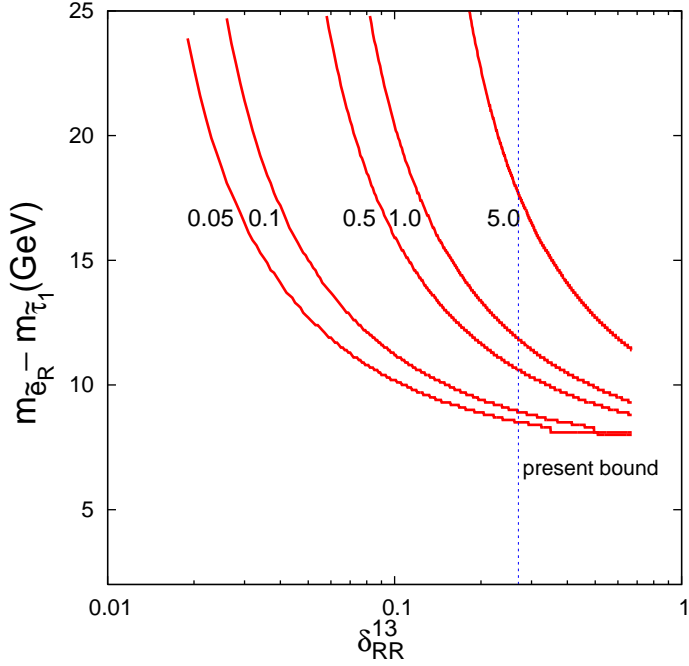}
\caption{\label{fig1} Contours of constant 
$\sigma(e^-e^- \rightarrow {\tilde e}_R
{\tilde \tau}^-_1 \rightarrow e^- \tau^+ {\tilde \tau}^-_1 {\tilde \tau}^-_1 
+ e^- \tau^- {\tilde \tau}^+_1 {\tilde \tau}^-_1$) in fb with $\sqrt{s}$ = 500 GeV, and the present experimental upper bound on $\delta^{13}$ coming
from the non-observation of the process $\tau\rightarrow e\gamma$.
The remaining parameters of the model are chosen as in
the $\epsilon$ point (see text for details).
Both $e^-$ beams are unpolarized.}
\end{figure}

\begin{figure}
\includegraphics{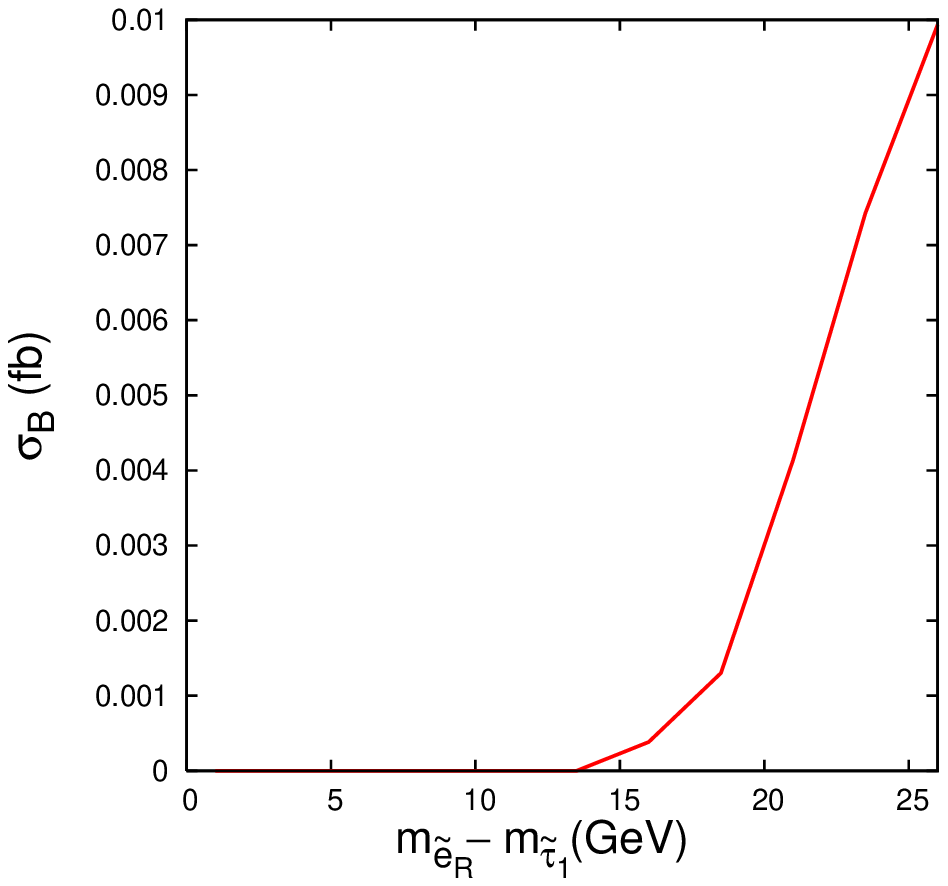}
\caption{\label{fig2} Upper bound on the background cross 
section (as discussed in the text)  $\sigma_B$ in fb in an $e^- e^-$ collider, 
with $\sqrt{s}$ = 500 GeV, as a function of the mass-difference between the
right-handed selectron and the NLSP. 
The remaining parameters of the model are chosen as in
the $\epsilon$ point (see text for details). Both  $e^-$ beams are 
unpolarized.}
\end{figure}

As discussed in previous sections, in the $e^-e^-$ mode
there are mainly two channels where lepton flavour violation
in the right-handed selectron-stau sector could be discovered,
namely processes with lepton flavour violating production
of sleptons:
\bea
e^-\; e^- &\rightarrow& \se^-_R\; \stau^-_1
\rightarrow (e^- \; \tau^{\pm} \; \stau^{\mp}_1)\;  \stau^-_1, 
\label{emem-seRstau1-2cl-III}
\eea
or with lepton flavour violating decay of selectrons:
\bea
e^-\; e^-&\rightarrow& \se^-_R\; \se^-_R \rightarrow 
(e^- \; e^{\pm} \; \stau^{\mp}_1)\; 
(e^- \; \tau^{\pm} \; \stau^{\mp}_1) ,
\label{emem-seRseR-3etau-III} \\
e^-\; e^-&\rightarrow& \se^-_R\; \se^-_R \rightarrow 
(\tau^- \; \tau^{\pm} \; \stau^{\mp}_1)\; 
(e^- \; \tau^{\pm} \; \stau^{\mp}_1).
\label{emem-seRseR-e3tau-III}
\eea
In Fig.\ref{fig1} we show contours of constant cross section for the 
process with lepton flavour violating production of sleptons, that
manifest itself as two heavily ionizing tracks and two charged leptons.
We show the results in the 
$\delta_{RR}^{13}$--$\Delta m$ plane with the cuts mentioned above 
(here, $\Delta m$ $\equiv$ $m_{{\tilde e}_R}$-- $m_{{\tilde \tau}_1}$.)
As discussed in the text, this process suffers from backgrounds 
stemming from the associated production of ${\tilde e}_L {\tilde e}_R$
and the subsequent decay of the left-handed selectron into neutrinos.
Nevertheless, imposing our cuts these background processes
have very small cross sections and can be neglected in general,
as can be realized from Fig. \ref{fig2}, where we show an upper
bound on the background cross section, calculated assuming for 
simplicity $BR({\tilde e}^-_L \rightarrow \nu_e {\bar \nu}_\tau 
{\tilde \tau}^-_1) = 1$. 

Demanding that the signal significance (defined as  $S/\sqrt{S+B}$ $\approx$
$\sqrt{S}$, where $S$ is the number of signal events and $B$ is the number of
background events) is larger than 5, we find that searching for this 
final state in the $e^-e^-$ mode, the
ILC could be sensitive to lepton flavour violation
down to the level $\delta^{13}_{RR}\sim 0.02$, when the supersymmetric
spectrum is as in the $\epsilon$ benchmark point ({\it i.e.} with
a mass splitting between  the right-handed selectron and the
NLSP of $\sim$ 20 GeV). The sensitivity would improve when the mass
splitting increases, whereas for  smaller
mass splittings, the outgoing charged leptons would be too soft and 
the cross section would be significantly reduced.  
For the  $\epsilon$ benchmark point, the sensitivity of the
ILC to lepton flavour violation is better than the present
sensitivity of experiments  searching for the rare decay 
$\tau\rightarrow e \gamma$, $\delta^{13}_{RR}\sim 0.27$ 
(cf. Table I) and comparable to the projected sensitivity of present
$B$-factories, $\delta^{13}_{RR}\sim 0.04$, that follows
from the projected bound $BR(\tau\rightarrow e\gamma)\lsim 10^{-8}$
\cite{Inami}
\footnote{The bounds on $\delta^{ij}_{RR}$ roughly scale
with $BR(l_j\rightarrow l_i \gamma)^{1/2}$.}.
On the other hand, future super-$B$ factories  could
produce of the order of $10^{10}$ $\tau$ pairs
at a luminosity of 10 $ab^{-1}$, allowing to probe
branching ratios  for the rare $\tau$ decays down
to the level of $10^{-8}-10^{-9}$ \cite{Akeroyd:2004mj}, 
which would translate into a sensitivity reach of
$\delta^{13}_{RR} \sim 0.01$. 

We have seen from the above discussion that the future sensitivity to
lepton flavour violation in rare decay experiments is comparable to that 
at the ILC in the $\epsilon$ benchmark point. However, there could be other
regions in the parameter space where the decay rate for $\tau\rightarrow e
\gamma$ is suppressed. This could be due to cancellations in the loops,
whereas the tree-level lepton flavour violating production of sleptons may
still remain unsuppressed. Furthermore, the observation of rare decays does
not shed any light on the source of lepton flavour violation: whether it
is in the left-handed sector, the right-handed sector or the trilinear
soft terms. On the other hand, the observation of the process $e^-\; e^-
\rightarrow (e^- \; \tau^{\pm} \; \stau^{\mp}_1)\;  \stau^-_1$ at the
linear collider would pinpoint the right-handed sector as one of the
sources of lepton flavour violation. Complementing this information with
the one from rare decays could help to identify the sources of flavour
violation in the leptonic sector. This could provide invaluable
information about the soft-breaking Lagrangian. To be more precise,
assuming that the  process $e^-\; e^-
\rightarrow (e^- \; \tau^{\pm} \; \stau^{\mp}_1)\;  \stau^-_1$
is observed at the ILC, the quantity $\delta^{13}_{RR}$
inferred from experiments could be used to predict
the rate for $\tau\rightarrow e\gamma$. If the observed rate
for this rare decay is larger than the predicted one,  it
would follow that there are necessarily additional sources of flavour
violation in the leptonic Lagrangian (either in $\delta^{13}_{LL}$,
  $\delta^{13}_{RL}$ or  $\delta^{13}_{LR}$). If these rates are
comparable, it would follow that the right-handed sector is the dominant
source of lepton flavour violation; and if the observed rate is smaller,
it would follow that different contributions to the decay amplitude are
canceling each other in order to produce a suppressed decay rate.

In this class of scenarios, lepton flavour violation 
could also be studied at future colliders using 
stopped staus \cite{stopped-staus}.
At the LHC, cascade decays of squarks and gluinos could
produce of the order of $10^6$ NLSPs per year if particle masses
are close to the current experimental limit. A fraction of them 
would be stopped in the walls of the detector,
\footnote{A larger number of staus could be trapped by placing 1-10 kton
massive material around the LHC detectors, to be precise around 
${\cal O}(10^3-10^4)$. Similarly, at the ILC
up to ${\cal O}(10^3-10^5)$ could be collected and studied.} and decay at late
times producing very energetic particles that would spring from
inside the detector. If there is no LFV, all the NLSPs would decay into
taus and gravitinos, $\stau\rightarrow \tau \psi_{3/2}$. If on the contrary
LFV exists in nature, some of the NLSPs would decay into electrons
and muons. Therefore the detection of very energetic particles
coming from inside the detector would constitute a signal of lepton flavour
violation. Using this technique, it has been estimated that 
at the LHC or the ILC it would be possible to probe down to 
$\delta^{13}_{RR} \sim 3\times 10^{-2}~ (9\times 10^{-3})$ if 
$3\times 10^3~(3\times 10^4)$ staus could be collected \cite{Hamaguchi:2004ne}.

\begin{figure}
\vspace*{-3.5in}
\includegraphics{eelfv-iacs.eps}
\includegraphics{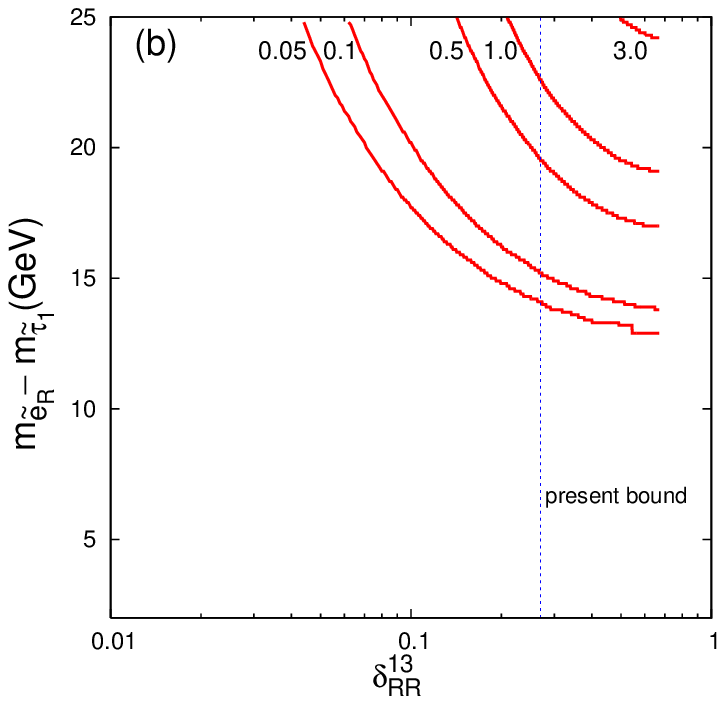}
\caption{\label{fig3} Contours of constant (a) $\sigma(e^-e^- \rightarrow {\tilde e}_R
{\tilde e}_R \rightarrow e^- e^- e^\pm \tau^\pm {\tilde \tau}^\mp_1 
{\tilde \tau}^\mp_1$) and (b) $\sigma(e^-e^- \rightarrow {\tilde e}_R
{\tilde e}_R \rightarrow \tau^- e^- \tau^\pm \tau^\pm {\tilde \tau}^\mp_1
{\tilde \tau}^\mp_1$) in fb with $\sqrt{s}$ = 500 GeV,
and the present experimental upper bound on $\delta^{13}$ coming
from the non-observation of the process $\tau\rightarrow e\gamma$.
The remaining parameters of the model are chosen as in
the $\epsilon$ point (see text for details).
Both  $e^-$ beams are unpolarized.}
\end{figure}

On the other hand, in Fig.\ref{fig3} we consider the lepton 
flavour violating signals in the decays
of the right-handed selectrons, Eqs.(\ref{emem-seRseR-3etau-III},
\ref{emem-seRseR-e3tau-III}), that can be detected as two 
heavily ionizing tracks and four charged leptons. We see from these
two figures that the lepton flavour violating signal from
$e^-\; e^-\rightarrow \se^-_R\; \se^-_R \rightarrow 
(e^- \; e^{\pm} \; \stau^{\mp}_1)\; 
(e^- \; \tau^{\pm} \; \stau^{\mp}_1)$
could be used to probe a larger region in the relevant 
parameter space compared to that from
$e^-\; e^-\rightarrow \se^-_R\; \se^-_R \rightarrow 
(\tau^- \; \tau^{\pm} \; \stau^{\mp}_1)\; 
(e^- \; \tau^{\pm} \; \stau^{\mp}_1)$. This is due to
the fact that the final states in these two processes are different and we have
used different $p_T$ cuts on the $\tau$s and the electrons (positron).

Although the source of lepton flavour violation is $\delta_{RR}^{13}$ in both 
Figs.(\ref{fig1}) and (\ref{fig3}), the region in the parameter space explored 
by the process in Fig.(\ref{fig1}) is larger than that in 
Figs.(\ref{fig3}(a)) or (\ref{fig3}(b)). There are two reasons for this feature.
In the case when the lepton flavour violation is in the decays of the right
selectron, due to the presence of ($m^2_{\widetilde l_R})_{13}$ there is a
branching ratio suppression in different channels shown in  Eq.(\ref{emem-seRseR-3etau})
and in Eq.(\ref{emem-seRseR-e3tau}). Also, since the final states are different
when one considers lepton flavour violation in production, the effects of the cuts
are also different.

\begin{figure}
\vspace*{-3in}
\includegraphics{mu-taulfv-iacs.eps}
\includegraphics{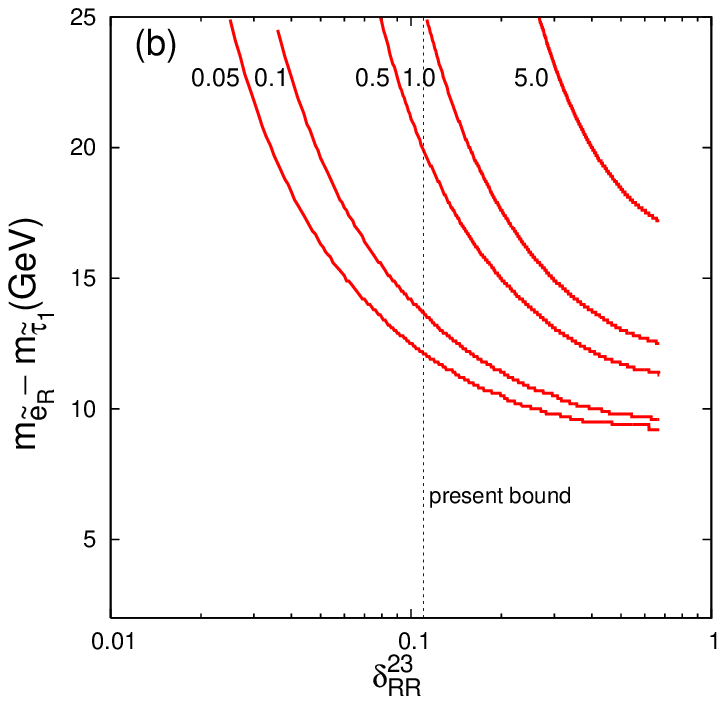}
\caption{\label{fig4} Contours of constant (a) $\sigma(e^-e^- \rightarrow {\tilde
e}_R {\tilde e}_R \rightarrow \mu^- e^- \tau^\pm \tau^\pm {\tilde \tau}^\mp_1
{\tilde \tau}^\mp_1$) and (b) $\sigma(e^-e^- \rightarrow {\tilde e}_R {\tilde e}_R
\rightarrow e^- e^- \mu^\pm \tau^\pm {\tilde \tau}^\mp_1 {\tilde \tau}^\mp_1$)
in fb with $\sqrt{s}$ = 500 GeV,
 and the present experimental upper bounds on $\delta^{12}$ and
 $\delta^{23}$  coming
from the non-observation of the processes $\mu\rightarrow e\gamma$ 
and $\tau\rightarrow \mu\gamma$, respectively. Note that the present
upper bound for $\delta^{12}$ lies outside figure (a).
The remaining parameters of the model are chosen as in
the $\epsilon$ point (see text for details).
Both  $e^-$ beams are unpolarized.}
\end{figure}

Other lepton flavour violating decays of the right-handed selectron 
which could be discovered in an $e^-e^-$ collider are:
\bea
e^-\; e^-&\rightarrow \se^-_R\; \se^-_R \rightarrow
(\mu^- \; \tau^{\pm} \; \stau^{\mp}_1)\;
(e^- \; \tau^{\pm} \; \stau^{\mp}_1)~~~~ {\rm if}~~ 
(m^2_{\widetilde l_R})_{12}\neq 0 ,
\label{emem-seRseR-mue2tau-III}\\
e^-\; e^-&\rightarrow \se^-_R\; \se^-_R \rightarrow
(e^- \; \mu^{\pm} \; \stau^{\mp}_1)\;
(e^- \; \tau^{\pm} \; \stau^{\mp}_1)~~~~ {\rm if}~~ 
(m^2_{\widetilde l_R})_{23}\neq 0.
\label{emem-seRseR-mu2etau-III}
\eea

In Figs.(\ref{fig4}a) and (\ref{fig4}b), we have shown contours of constant 
cross sections of these two processes in the plane ($\delta_{RR}^{12}$ --$\Delta m$)
for the process in Eq.(\ref{emem-seRseR-mue2tau-III}) and similarly in the plane
($\delta_{RR}^{23}$ --$\Delta m$) for the process in 
Eq.(\ref{emem-seRseR-mu2etau-III}). Obviously, the reach in the process in
Eq.(\ref{emem-seRseR-mu2etau-III}) is larger since it contains only one $\tau$ and
hence is less affected by the $p_T$ cuts. On the other hand, if one compares the
process in Eq.(\ref{emem-seRseR-3etau-III}) with the process in
Eq.(\ref{emem-seRseR-mu2etau-III}) then one can observe that though these two 
processes look very similar (since we impose similar $p_T$ cuts for the 
$e^\pm$ and the $\mu^\pm$), the behaviours are different for higher values of 
the corresponding parameters $\delta_{RR}^{ij}$ and $\Delta m$. This is again due 
to the fact that there is a branching ratio suppression in the process in 
Eq.(\ref{emem-seRseR-3etau-III}) for relatively large values of the 
$\delta_{RR}^{13}$ and $\Delta m$. For very low values of $\delta_{RR}^{ij}$ the 
nature is very similar in both the figures. Again, for low values of $\Delta m$ 
and large values of $\delta_{RR}^{ij}$ these two figures show quite similar 
behaviour since the branching ratio suppression in the process in 
Eq.(\ref{emem-seRseR-3etau-III}) is less pronounced in this region and the lower
limits in $\Delta m$ are determined by the $p_T$ cuts employed.

As before, if we demand that the signal significance is greater than or
equal to 5, for the $\epsilon$ benchmark point 
the ILC could probe lepton flavour violation down to the level 
$\delta^{12}_{RR}\sim 0.04$, $\delta^{23}_{RR}\sim 0.03$,
{\it i.e.} for a mass splitting
between the right-handed selectron and the NLSP of $\sim 20$ GeV.
For this benchmark point,
the sensitivity to lepton flavour violation in the smuon-stau sector is 
slightly better than the present sensitivity from the rare decay 
$\tau\rightarrow \mu \gamma$, and in the smuon-selectron sector,
is much worse than from the decay $\mu\rightarrow e\gamma$ 
(cf. Table I).

For the sake of completeness, let us now discuss the process
$e^-\; e^- \rightarrow \se^-_R\; \smu^-_R\rightarrow
(e^- \; \tau^{\pm} \; \stau^{\mp}_1) 
(\mu^- \; \tau^{\pm} \; \stau^{\mp}_1)$. 
The cross section for this process should be 
slightly smaller than for the process 
$e^-\; e^- \rightarrow \se^-_R\; \stau^-_1
\rightarrow (e^- \; \tau^{\pm} \; \stau^{\mp}_1)\;  \stau^-_1$,
due to the larger multiplicity of the final state of the former
process, that translates into a bigger impact of the kinematical
cuts and the reduction of the signal strength. Thus, the 
sensitivity to $\delta^{12}_{RR}$ through the observation of the process 
with associated production of a right-handed selectron and a right-handed
smuon, Eq.(\ref{emem-seRsmuR-mue2tau}), should be slightly smaller 
than the sensitivity to $\delta^{13}_{RR}$ coming from the associated 
production of a right-handed selectron and a NLSP,
Eq.(\ref{emem-seRstau1-2cl}). On the 
other hand, the final state for the process with associated production 
of a right-handed selectron and a right-handed smuon,
Eq.(\ref{emem-seRsmuR-mue2tau}), is the same as 
the final state for the pair production of two right-handed selectrons,
followed by the flavour violating decay of one of them into a muon, a tau
and a NLSP, Eq.(\ref{emem-seRseR-mue2tau}). 
If we combine the cross sections of 
these two processes to search for lepton 
flavour violation, the sensitivity to $\delta^{12}_{RR}$ will increase 
significantly. 

From the above discussion we can see that different regions 
in the corresponding parameter space can be explored in the lepton 
flavour violating production and decays
of the right-selectron in a future $e^-e^-$ collider.

\subsection{$e^+\; e^-$ collider}

Next, we will discuss the situation in an $e^+e^-$ collider. As discussed in Section
IIB, lepton flavour violating production processes are similar to the ones in the
$e^-e^-$ collider with the corresponding changes in the electric charge of the
particles. We would also like to reiterate that only the $t$-channel neutralino
mediated diagrams are taken into account for the calculation of these production
processes, since the $s$-channel contributions through photon and 
$Z$-boson exchange are
loop-suppressed. Let us first consider the case when only $\delta^{13}_{RR}$ is
non-vanishing. In this scenario the lepton flavour violating 
production process in an $e^+e^-$ collider looks like
\bea
e^+\; e^-&\rightarrow& \se^{+}_R\; \stau^{-}_1 + \se^{-}_R\; \stau^{+}_1 \rightarrow
(e^+ \;\tau^{\pm} \;\stau^{\mp}_1)\; \stau^{-}_1 + (e^- \;\tau^{\pm}
\;\stau^{\mp}_1)\; \stau^{+}_1. 
\label{epem-seRstau1-2cl-III}
\eea 

\begin{figure}
\includegraphics{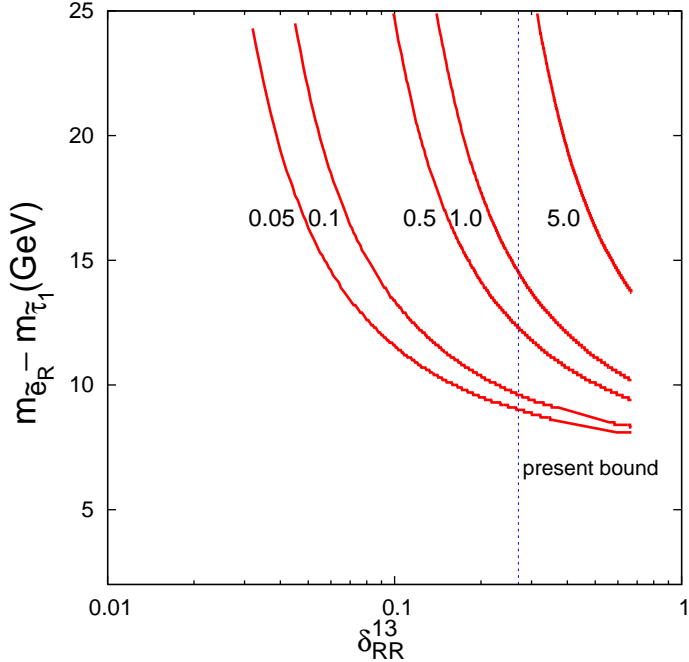}
\caption{\label{fig5} Contours of constant cross sections $\sigma(e^+e^- \rightarrow
{\tilde e}^+_R {\tilde \tau}^-_1 + {\tilde e}^-_R {\tilde \tau}^+_1 \rightarrow
e^+ \tau^\mp {\tilde \tau}^\pm_1 {\tilde \tau}^-_1 + e^- \tau^\mp {\tilde \tau}^\pm_1
{\tilde \tau}^+_1$) in fb with $\sqrt{s}$ = 500 GeV,
and the present experimental upper bound on $\delta^{13}$ coming
from the non-observation of the process $\tau\rightarrow e\gamma$.
The remaining parameters of the model are chosen as in
the $\epsilon$ point (see text for details).
Both the $e^+$ and the $e^-$ beam are unpolarized.}
\end{figure}

\begin{figure}
\includegraphics{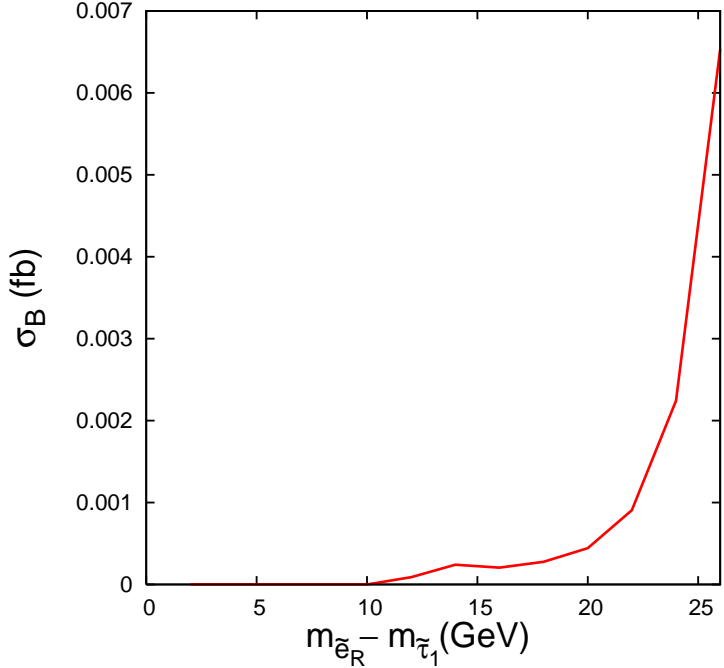}
\caption{\label{fig6} Upper bound on the background cross section
 (as discussed in the text)  $\sigma_B$ in fb in a $e^+ e^-$ collider, 
with $\sqrt{s}$ = 500 GeV, as a function of the 
mass-difference between the right-handed selectron and the NLSP.
The remaining parameters of the model are chosen as in
the $\epsilon$ point (see text for details). Both the $e^+$ 
and $e^-$ beams are unpolarized.}
\end{figure}

The contours of constant cross sections of the process in
Eq.(\ref{epem-seRstau1-2cl-III}) have been plotted in Fig.\ref{fig5} in the 
($\delta_{RR}^{13}$ -- $\Delta m$) plane for other parameter choices as in
Fig.\ref{fig1} and with $\sqrt s$ = 500 GeV. The background to this process 
can come from the associated production $e^+e^- \rightarrow \se^\pm_R\;
\se^\mp_L$ as shown in Eq.(\ref{e+e-lfvprod-bkgd}) (with the corresponding 
modifications for the charge conjugate process). The cross section of the 
background process is plotted in Fig.\ref{fig6} as a function of the mass 
difference between the right-selectron and the NLSP. Comparing Fig.\ref{fig1} 
and Fig.\ref{fig5} and looking at the background cross section we observe 
that the parameter region which can be explored in an $e^+e^-$ collider 
with 5$\sigma$ significance is slightly smaller than in the case of an 
$e^-e^-$ collider, to be precise, $\delta_{RR}^{13}\gsim0.03$ against
$\delta_{RR}^{13}\gsim0.02$ for $\Delta m\lsim 20$ GeV.
This is due to the fact that the production cross section 
of the lepton flavour violating process in an $e^+e^-$ collider is smaller 
compared to the corresponding process in an $e^-e^-$ collider. On the other 
hand, the smaller cross sections in the $e^+ e^-$ for the lepton flavour 
violating processes could be compensated with a bigger luminosity, as will 
presumably occur since the electrons in the $e^-e^-$ collider repel each 
other translating into a decrease in the luminosity with respect to the 
$e^+e^-$ mode. Also, using right-polarized electron and positron
beams the signal strength could be enhanced and the background cross 
section can be reduced further. However, as discussed earlier the sensitivity 
to lepton flavour violation does not change significantly because of the 
reduction in luminosity for polarized beams.

\begin{figure}
\vspace*{-3.5in}
\includegraphics{epem_eelfv_iacs.eps}
\includegraphics{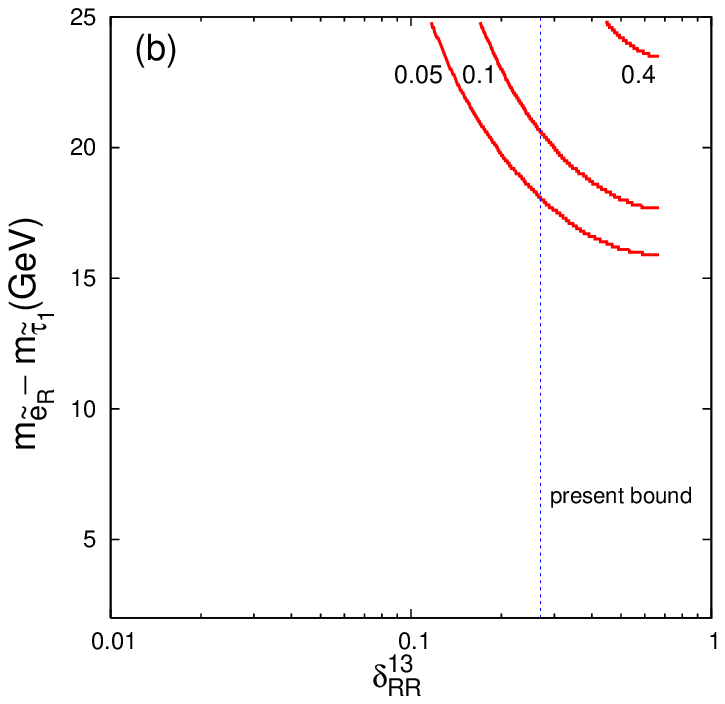}
\caption{\label{fig7} Contours of constant (a) $\sigma(e^+e^- \rightarrow 
{\tilde e}^+_R {\tilde e}^-_R \rightarrow e^+ e^- e^\pm \tau^\pm 
{\tilde \tau}^\mp_1 {\tilde \tau}^\mp_1$) and (b) $\sigma(e^+e^- 
\rightarrow {\tilde e}^+_R {\tilde e}^-_R \rightarrow \tau^+ e^- \tau^\pm 
\tau^\pm {\tilde \tau}^\mp_1 {\tilde \tau}^\mp_1 + \tau^- e^+ \tau^\pm
\tau^\pm {\tilde \tau}^\mp_1 {\tilde \tau}^\mp_1$) in fb with $\sqrt{s}$ 
= 500 GeV, 
and the present experimental upper bound on $\delta^{13}$ coming
from the non-observation of the process $\tau\rightarrow e\gamma$.
The remaining parameters of the model are chosen as in
the $\epsilon$ point (see text for details).
Both the $e^+$ and the $e^-$ beam are unpolarized.}
\end{figure}

In Fig.\ref{fig7}, we consider lepton flavour violation in 
right-handed selectron decays as in the case of an $e^-e^-$
 collider. These signals 
are generated from the ${\tilde e}^+_R {\tilde e}^-_R$ pair production 
followed by their lepton flavour violating decays and can be detected as 
four charged leptons and two heavily ionizing charged tracks. Looking at
Fig.\ref{fig7}(b) and Fig.\ref{fig5}, we see that the process 
$e^+\; e^-\rightarrow \se^+_R\; \se^-_R \rightarrow 
(\tau^+ \; \tau^{\pm} \; \stau^{\mp}_1)\; 
(e^- \; \tau^{\pm} \; \stau^{\mp}_1)$
is not an efficient way to search for lepton flavour violation, compared
to the process with two charged leptons in the final state. On the other 
hand, Fig.\ref{fig7}(a) shows that the sensitivity of the $e^+e^-$ collider 
for the lepton flavour violating process 
$e^+\; e^-\rightarrow \se^+_R\; \se^-_R \rightarrow 
(e^+ \; e^{\pm} \; \stau^{\mp}_1)\; 
(e^- \; \tau^{\pm} \; \stau^{\mp}_1)$ is better than 
the present sensitivity of the experiments, $\delta^{13}_{RR} 
\sim$ 0.27 when $\Delta m \sim$ 20 GeV, although not as good
as the corresponding sensitivity in an $e^-e^-$ collider (see 
Fig.\ref{fig3}). This is again because of the fact that the right-selectron 
pair production cross section is larger in an $e^-e^-$ collider since both 
$t$- and $u$-channel diagrams are present and they interfere constructively.

\begin{figure}
\vspace*{-3.5in}
\includegraphics{epem_emlfv_iacs.eps}
\includegraphics{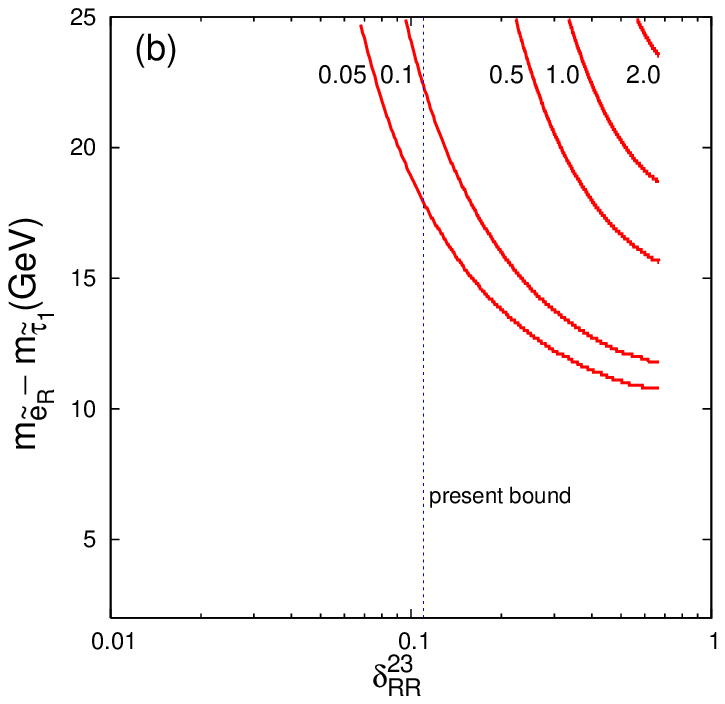}
\caption{\label{fig8} Contours of constant (a) $\sigma(e^+e^- \rightarrow 
{\tilde e}^+_R {\tilde e}^-_R \rightarrow \mu^+ e^- \tau^\pm \tau^\pm 
{\tilde \tau}^\mp_1 {\tilde \tau}^\mp_1 + \mu^- e^+ \tau^\pm \tau^\pm
{\tilde \tau}^\mp_1 {\tilde \tau}^\mp_1$) and (b) $\sigma(e^+e^- 
\rightarrow {\tilde e}^+_R {\tilde e}^-_R \rightarrow e^+ e^- \mu^\pm 
\tau^\pm {\tilde \tau}^\mp_1 {\tilde \tau}^\mp_1$) in fb with $\sqrt{s}$ 
= 500 GeV, 
 and the present experimental upper bounds on $\delta^{12}$ and
 $\delta^{23}$  coming
from the non-observation of the processes $\mu\rightarrow e\gamma$ 
and $\tau\rightarrow \mu\gamma$, respectively. Note that the present
upper bound for $\delta^{12}$ lies outside figure (a).
The remaining parameters of the model are chosen as in
the $\epsilon$ point (see text for details).
Both the $e^+$ and the $e^-$ beams are unpolarized.}
\end{figure}

In order to find out the sensitivity to $\delta^{12}_{RR}$ and
$\delta^{23}_{RR}$ in an $e^+e^-$ collider, we have plotted in Fig.
\ref{fig8} the contours of constant cross sections for the processes
\bea
e^+\; e^-&\rightarrow \se^+_R\; \se^-_R \rightarrow 
(\mu^+ \; \tau^{\pm} \; \stau^{\mp}_1)\; 
(e^- \; \tau^{\pm} \; \stau^{\mp}_1),
\label{epem-seRseR-mue2tau-copy}\\
e^+\; e^-&\rightarrow \se^+_R\; \se^-_R \rightarrow 
(e^+ \; \mu^{\pm} \; \stau^{\mp}_1)\; 
(e^- \; \tau^{\pm} \; \stau^{\mp}_1),
\label{epem-seRseR-mu2etau-copy}
\eea
respectively.
Demanding that the signal significance is greater or
equal to 5, we obtain that the ILC in the $e^+e^-$ mode
could be sensitive to $\delta^{12}_{RR}\sim 0.1$
and $\delta^{23}_{RR}\sim 0.08$  when $\Delta m\sim20$ GeV.
Once again, the sensitivity to these flavour violating 
quantities are slightly poorer than  at the $e^-e^-$ collider.   

As we also argued in the case of $e^-e^-$ collider, the process 
$e^+\; e^-\rightarrow  \se^{+}_R \; \smu^{-}_R \rightarrow 
(e^+ \;\tau^{\pm} \;\stau^{\mp}_1)\;  (\mu^- \;\tau^{\pm} \;\stau^{\mp}_1)$
should have a cross section slightly smaller than the process
$e^+\; e^-\rightarrow \se^{+}_R\; \stau^{-}_1 \rightarrow 
(e^+ \;\tau^{\pm} \;\stau^{\mp}_1)\; \stau^{-}_1$ 
and hence the sensitivity to $\delta^{12}_{RR}$ 
through the observation of the former process is slightly poorer than
the  sensitivity to $\delta^{13}_{RR}$ obtained from the observation of the
latter. However, as pointed out in the $e^-e^-$ 
case, the final state of the process with associated production
of a right-handed selectron and a right-handed smuon is identical
to that in Eq.(\ref{epem-seRseR-mue2tau-copy})
 and combining the two cross sections should 
improve the sensitivity of $\delta^{12}_{RR}$ to a significant extent. 

Finally, let us discuss briefly the processes with pair production
of two right-handed smuons, followed by the lepton flavour violating
decay of one of them, Eqs.(\ref{epem-smuRsmuR-3mutau})--(\ref{epem-smuRsmuR-e2mutau}).
The production cross section of a pair of right-handed smuons
is suppressed approximately by a factor of 5 compared to the 
right-handed selectron pair production. In consequence, the constraints
on the different $\delta^{ij}_{RR}$ from the non-observation of these
processes are approximately a factor of five weaker than the constraints
coming from processes with pair production of right-handed selectrons,
and are therefore poorer probes of lepton flavour violation.

\section{Conclusions}

In this paper we have estimated the sensitivity of future $e^+e^-$ and $e^-e^-$ 
colliders to lepton flavour violation, in scenarios where the gravitino is
the lightest supersymmetric particle (LSP) and the stau is the next-to-lightest
supersymmetric particle (NLSP). Since the NLSP can only decay gravitationally
into gravitinos and charged leptons, the decay rate is very suppressed and the 
NLSP could traverse several layers of the vertex detector before decaying or 
even being stopped and trapped in it. This peculiar signature would be a clear 
signal for this class of scenarios and in particular would allow the clean 
search for lepton flavour violation, as the Standard Model backgrounds are 
very small and the supersymmetric backgrounds can be kept under control by 
using suitable kinematic cuts.

The signals of lepton flavour violation would consist of two heavily 
ionizing tracks due to the long-lived staus accompanied by two or four 
charged leptons.   
Final states with two heavily ionizing tracks and two charged leptons 
correspond to the lepton flavour violating production 
of a right-handed selectron and a NLSP, followed by the decay of the 
right-handed 
selectron into the NLSP and two charged leptons, and would constitute a
signal for lepton flavour violation in the right-handed selectron-stau sector.
On the other hand, final states with  two heavily ionizing tracks and 
four charged leptons correspond to the pair production of two
right-handed selectrons (and also smuons, in the case of the $e^+e^-$ 
collider), 
followed by one lepton flavour violating decay and one lepton flavour 
conserving
decay into the NLSP and two charged leptons. These signals arise when there
exists mixing between any two generations of the right-handed sector.
Nevertheless, to search for lepton flavour violation
in the selectron-stau sector, we find that signals with two  charged 
leptons in the final state are a more sensitive probe 
than signals with four charged leptons. We also find
that the sensitivity to lepton flavour violation is slightly better at the 
$e^-e^-$ collider than at the  $e^+e^-$ collider, due to the slightly
larger production cross-section of sleptons at the $e^-e^-$ collider (either in 
a lepton flavour conserving or in a lepton flavour violating mode), which is 
due to the  constructive interference of the $t$- and the $u$-channel 
production amplitudes.

To illustrate the sensitivity reach of this experiment, we have analyzed in 
detail a variant of the $\epsilon$ benchmark point presented in 
\cite{DeRoeck:2005bw}, taking all the supersymmetric parameters as in the 
$\epsilon$ benchmark point, but varying the NLSP mass between 144 GeV and 
167 GeV and admitting some small amount
of lepton flavour violation in the right-handed slepton sector,
parametrized by $\delta_{RR}^{ij}$. We have also estimated the efficiency 
of detecting the long-lived staus using the traditional methods and folded the 
efficiency with the calculated signal cross sections.

In particular, when
the mass splitting between the NLSP and the next-to-NLSP is ~20 GeV,
we find that the International Linear Collider with a center of mass energy
of $\sqrt{s}=500$ GeV and an integrated luminosity of 500 fb$^{-1}$ could
probe lepton flavour violation down to the level $\delta_{RR}^{13}\sim0.02$,
 $\delta_{RR}^{12}\sim0.04$, $\delta_{RR}^{23}\sim0.03$ at 5$\sigma$
in the $e^-e^-$ mode using unpolarized beams,
 and slightly worse in the $e^+e^-$ mode.
As a side remark we would like to mention that the use of right-polarized 
electron and positron beams enhances the signal strength and reduces 
the background cross section. However, the sensitivity to lepton flavour 
violation does not improve significantly since the integrated luminosity
is also reduced for polarized beams.
This sensitivity is competitive with the present and projected sensitivities
to lepton flavour violation from rare tau decays, although not from rare muon 
decays where the non-observation of the process $\mu\rightarrow e\gamma$ still 
gives the most stringent constraints. Finally, we would like to remark that 
whereas the origin of the lepton flavour violation cannot be determined just 
by the observation of rare decays, the observation of the tree level production and/or decay of sleptons
at the International Linear Collider would pinpoint the right-handed slepton
sector as one of the sources of lepton flavour violation.
Complementing this information with the one from rare decays
could help to identify the sources of flavour violation
in the leptonic sector, providing invaluable information
about the soft-breaking Lagrangian.

\section*{Appendix}

We review here the relevant formulas to compute the bounds
on the lepton flavour violating parameters $\delta_{LL}$,
$\delta_{RR}$, $\delta_{LR}$, $\delta_{RL}$ 
from the non-observation of the rare decays $\ell_i\rightarrow \ell_j \gamma$,
following closely the analysis by Masina and Savoy \cite{Masina:2002mv}.
In the mass insertion approximation, the branching ratio for the
process $\ell_i\rightarrow \ell_j \gamma$ reads:
\bea
\label{BR-theoretical}
&&{\rm BR}(\ell_i \rightarrow \ell_j \gamma) = 3.4\times 10^{-4}~
{\rm BR}(\ell_i \rightarrow \ell_j \bar\nu_j \nu_i)~ 
\frac{M_W^4 M_1^2 \tan^2\beta}{|\mu |^2}
\times \nonumber \\
&&  \left\{
\left|\ \delta^{LL}_{ji} ( \eta^* I'_{B,L} + \frac{1}{2} I'_L + I'_2 )  
+ \delta^{LR}_{ji}\frac{m^{av}_R m^{av}_L}{\mu~ m_i \tan\beta}I_B \right|^2 + 
\left|\delta^{RR}_{ji}( \eta I'_{B,R} - I'_R ) + 
\delta^{RL}_{ji}\frac{m^{av}_R m^{av}_L}{\mu^* m_i \tan\beta} I_B \right|^2
\right\}\;, \nonumber \\
\eea
where we have defined
\bea
\eta&=&1-\frac{A^{av}_l}{\mu^*\tan\beta} \;,\nonumber \\
I_B&=&\frac{1}{{m^{av}_{R}}^2 - {m^{av}_{L}}^2} 
\left[ \frac{|\mu|^2}{{m^{av}_L}^2}  g_1 \left(\frac{M_1^2}{{m^{av}_L}^2} \right) 
-  \frac{|\mu|^2}{{m^{av}_R}^2}  g_1 \left(\frac{M_1^2}{{m^{av}_R}^2}\right) \right]\;,
\nonumber \\
I'_L &=& \frac{1}{{m^{av}_L}^2}
\frac{|\mu|^2}{|\mu|^2-M_1^2}
\left[ h_1 \left( \frac{M_1^2}{{m^{av}_L}^2}\right)
- \, h_1 \left( \frac{|\mu|^2}{{m^{av}_L}^2} \right)\right]\;,
\nonumber \\
I'_R &=& \frac{1}{{m^{av}_R}^2}
\frac{|\mu|^2}{|\mu|^2-M_1^2}
\left[ h_1 \left( \frac{M_1^2}{{m^{av}_R}^2}\right)
- \, h_1 \left( \frac{|\mu|^2}{{m^{av}_R}^2} \right)\right]\;,
\nonumber \\
I'_2 &=& \frac{M_2\cot ^2\theta_W}{M_1 {m^{av}_L}^2}
\frac{|\mu|^2}{|\mu|^2-|M_2|^2}
\left[\, h_2 \left( \frac{|M_2|^2}{{m^{av}_L}^2}\right) 
- \, h_2 \left( \frac{|\mu|^2}{{m^{av}_L}^2}\right)\right]\;,
\nonumber \\
I'_{B,R}&=&  -\frac{1}{{m^{av}_R}^2 - {m^{av}_L}^2}
\left[  \frac{|\mu|^2}{{m^{av}_R}^2}  h_1\left(\frac{M_1^2}{{m^{av}_R}^2}\right)  
- {m^{av}_R}^2 I_B \right]\;,
\nonumber \\
I'_{B,L}&=&  -\frac{1}{{m^{av}_L}^2 - {m^{av}_R}^2}
\left[  \frac{|\mu|^2}{{m^{av}_L}^2}  h_1\left(\frac{M_1^2}{{m^{av}_L}^2}\right)  
- {m^{av}_L}^2 I_B \right]\;.
\eea
The functions $g_{1,2}$ and $h_{1,2}$ have the following expression:
\bea
g_1 (x)&=& \frac{1-x^2+2x\ln (x)}{(1-x)^3}\;,  \nonumber \\
h_1 (x)&=& \frac{1+4x-5x^2+(2x^2+4x)\ln(x)}{(1-x)^4}\;, \nonumber  \\
h_2 (x)&=& \frac{7x^2+4x-11-2(x^2 +6x+2)\ln(x)}{2(x-1)^4} \;.
\eea
In these formulas, $m^{av}_L$, $m^{av}_R$, $A^{av}_l$ were defined after 
eq.(\ref{mass-matrix2})\footnote{Note that in contrast
to \cite{Masina:2002mv} we have absorbed all the flavour
dependence of $A_{l_{ij}}$ in the definitions of $\delta_{LR,RL}$.},
$m_i$ is the mass of the decaying lepton, $\theta_W$ is the Weinberg's
angle, $M_1$ is the bino mass (that we choose real, by means of a phase
redefinition) and $M_2$ is the wino mass. 
The bounds on the $\delta$ parameters in table \ref{Table1}
can be straightforwardly computed using eq.(\ref{BR-theoretical}),
assuming that one single $\delta$ is the only source
of lepton flavour violation.

\section{Acknowledgments}

We thank Emidio Gabrielli, Dilip Kumar Ghosh, Koichi Hamaguchi, 
Andrea Romanino, 
Xerxes Tata, and Sudhir Vempati for very helpful discussions. 
We would also like to thank the anonymous referee of JHEP for
his/her useful suggestions. SR thanks the 
ASICTP for kind hospitality during the preparation of this work.


\end{document}